\documentclass[aps,prc,twocolumn,groupedaddress,nofootinbib]{revtex4}

\usepackage{graphicx}% Include figure files
\newcommand{\si}{\sigma}

\newcommand{\oeq}{\begin{equation}}
\newcommand{\ceq}{\end{equation}}
\newcommand{\oeqn}{\begin{eqnarray}}
\newcommand{\ceqn}{\end{eqnarray}}

\newcommand{\sdf}{\,\,}

\begin{document}

\title{Microscopic description of heavy ion collisions around the barrier}

\author{C\'edric Simenel and Beno\^{ii}t Avez}
 \affiliation{
CEA, Centre de Saclay, IRFU/Service de Physique Nucl\'eaire, F-91191 Gif-sur-Yvette, France.}

\author{C\'edric Golabek}
\affiliation{
GANIL (IN2P3/CNRS - DSM/CEA), BP 55027, F-14076 Caen Cedex 5, France.
}

\date{\today}

\begin{abstract}
A microscopic mean-field description of heavy ion collisions
is performed in the framework of the time dependent Hartree-Fock 
 theory using a Skyrme energy density functional.
 A good agreement with experiments is obtained on the position of the fusion barriers 
for various total masses and mass asymmetries.
The excitation function of the $^{16}$O+$^{208}$Pb is overestimated by about
16$\%$ above the barrier. 
Transfer below the barrier is studied in $^{16}$O+$^{208}$Pb central collisions. 
Heavier systems are considered to study their fusion hindrance.
We also compute collision times of the $^{238}$U+$^{238}$U system.
The latter has been used to produce super-strong electric fields and to test non perturbative quantum electrodynamics theory. Indeed, if the life time of such giant system is of the order of few $10^{-21}$~s, its electric field 
should induce spontaneous electron-positron pair emissions from vacuum.
In our calculations, highest collision times are reached in the $^{238}$U+$^{238}$U reaction for center of mass energies between 1000 and 1300 MeV.
\end{abstract}


\maketitle

\section{introduction}

Description of nuclear reactions is very challenging, 
especially at energies around the fusion barrier
generated by the competition between Coulomb and nuclear
interactions.
It has been established that the structure of the collision partners
may affect strongly the reaction mechanisms in this energy domain, as,
for instance, the fusion cross-sections (for a review, see, {e.g.}, Ref.~\cite{das98}).
In fact, a good understanding of these effects is crucial for the quest of superheavy elements (SHE)~\cite{hof00}. Indeed, different structures of the collision partners may change SHE production cross sections by several orders of magnitudes~\cite{hin08}.
For instance, fusion hindrance has been observed for collisions of quasisymmetric heavy systems with typical charge products $Z_1\times Z_2\geq 1600$~\cite{gag84}.
Here, dynamical shape evolution of the dinuclear system and transfer between fragments are expected to play a significant role on this fusion hindrance. %Ref ?
Investigations on collision dynamics between actinides are also of interest, in particular to study the physics of superstrong electric fields~\cite{rei81,gre83,ack08}.

It is therefore recommended to treat both structure of nuclei and their collision dynamics 
within the same formalism. This is the case of fully microscopic approaches 
such as the time dependent Hartree-Fock (TDHF) theory 
proposed by Dirac in 1930~\cite{dir30}. This time dependent version
of the well known  Hartree-Fock (HF) theory~\cite{har28,foc30}
gives a self-consistent mean-field description of nuclear dynamics. 
The success of the first HF calculations based on Skyrme interactions~\cite{sky56,vau72} led to tremendous activities to describe
 nuclear structure within mean-field based approaches 
(see, e.g., Ref.~\cite{ben03} for a review). 
Early TDHF calculations in nuclear physics~\cite{eng75,bon76} have been done with the seek for a description of the dynamics of nuclei as good as their static 
properties~\cite{neg82}.
At that time, theoreticians had to use various symmetries and 
simplified Skyrme interactions to reduce computational 
times. Now, recent increase of computational power allowed 
realistic TDHF calculations of nuclear collisions in 3 dimensions 
with full Skyrme energy density functional 
(EDF)~\cite{kim97,uma06a}.

In this proceeding, we present a review of various TDHF studies of heavy ion collisions.
In Section~\ref{sec:TDHF}, we recall the formalism and detail the calculation.
In Section~\ref{sec:fus_light}, we study fusion and transfer in light and medium heavy
systems. In particular, a detailed study of the $^{16}$O+$^{208}$Pb reaction is presented.
In Section~\ref{sec:fus_heavy}, we investigate the problem of fusion hindrance of heavy quasisymmetric systems. Then, we study the collision time of two $^{238}$U nuclei
in Section~\ref{sec:U+U} before to conclude.

\section{The Time-Dependent Hartree-Fock Approach}
\label{sec:TDHF}

In nuclear physics, the TDHF theory is applied with a Skyrme EDF modeling nuclear interactions between nucleons~\cite{sky56}. The EDF is the only phenomenological ingredient of the model, as it has been adjusted on nuclear structure properties like infinite nuclear matter and radii and masses of few doubly magic nuclei~\cite{cha98}. The main approximation of the theory is to constrain the many-body wave function to be an antisymmetrized independent particles state at any time. The latter ensures an exact treatment of the Pauli principle during time evolution. Though TDHF does not include two-body collision term, it is expected to treat correctly one-body dissipation which is known to drive low energy reaction mechanisms as Pauli blocking prevents nucleon-nucleon collisions.

The main advantage of TDHF is that it treats static
properties {\it and} dynamics of nuclei within the same
formalism and the same EDF. The initial state is obtained
through static HF calculations which 
give a good approximation of nuclear binding energies and
deformations. 
Another important advantage of TDHF for near-barrier reaction studies 
is that it contains all  types of couplings between the relative
motion and internal degrees of freedom whereas in coupled
channels calculations one has to include them explicitly
according to physical intuition. 
However, TDHF gives only classical trajectories for the time-evolution  and expectation values of one-body observables. In particular, it does not include tunneling of the many-body wave function.

Inclusion of pairing correlations responsible for superfluidity in nuclei have been done  recently within the time dependent Hartree-Fock-Bogolyubov theory to study pairing vibrations in nuclei~\cite{ave08}. However, realistic applications to heavy ion collisions are not yet achieved and are beyond the scope of this work.

\subsection{Formalism}

The TDHF equation can be written as a Liouville-Von Neumann equation 
\begin{equation}
i\hbar \frac{\partial}{\partial t} \rho = \left[h[\rho],\rho\right]
\label{eq:tdhf}
\end{equation}
where $\rho$ is the one body density matrix associated to the total independent particles state with elements 
\begin{equation}
\rho(\mathbf{r} sq, \mathbf{r'}s'q') = \sum_{i=1}^{A_1+A_2} \sdf  \varphi_i(\mathbf{r} sq)\sdf \varphi_i^*(\mathbf{r'}s'q').
\end{equation}
The sum runs over all occupied single particle wave functions $\varphi_i$ and $\mathbf{r}$, $s$ and $q$ denote the position, spin and isospin of the nucleon respectively.
The Hartree-Fock single particle Hamiltonian $h[\rho]$ is related to the EDF, noted $E[\rho]$, by its first derivative
%\begin{equation}
$h[\rho](\mathbf{r} sq, \mathbf{r'}s'q') = \frac{\delta E[\rho]}{\delta \rho(\mathbf{r'} s'q', \mathbf{r} sq)}$.
%\end{equation}

\subsection{Practical aspects}

A TDHF calculation of two colliding nuclei is performed assuming that 
the two collision partners are initially at a distance
$D_0$ in their HF ground state. This distance has to be big enough 
to allow Coulomb excitation in the entrance channel 
(polarization, vibration, rotation...).
We chose $D_0=44.8$~fm in sections~\ref{sec:fus_light} and~\ref{sec:fus_heavy} and $D_0=51.2$~fm in sections~\ref{sec:U+U}. 
We assume that before to reach this distance,
the nuclei followed a Rutherford trajectory which determines their 
initial velocities. A Galilean transformation~\cite{tho62} is then applied on each nucleus 
in the first iteration.

We use the  {\textsc{tdhf3d}} code built by P. Bonche and coworkers with the SLy4$d$
 Skyrme parametrization~\cite{kim97}.
 This code has been extensively used to study heavy ion fusion~\cite{sim01,sim04,sim07,sim08,sim08b,was08}. 
It has a plane of symmetry (the collision plane).
It uses the Skyrme energy functional expressed in Eq.~(A.2) of Ref.~\cite{bon87} 
where the tensor coupling between spin and gradient has been neglected.
The lattice spacing is $\Delta x=0.8$~fm and the time step is $\Delta t=1.5\times10^{-24}$~s.

\section{Fusion and transfer in light and medium-heavy systems}
\label{sec:fus_light}

In this section, we focus on reactions with spherical nuclei. 
Studies of the effects of deformation on fusion within TDHF can be found in~\cite{sim04,uma06b,sim08,sim08b}.

\subsection{Fusion barriers}

Fusion barriers are classically defined as the energy threshold above which  
fusion occurs for a head-on collision. Experimentally, 
the average position of the barrier can be approximated by the centroid 
of the so-called barrier distribution~\cite{row91}. 
To determine fusion barriers from TDHF, we consider head-on collisions
at various energies. The barrier of a system is then located between the highest energy 
for which there is no fusion and the lowest one for which fusion occurs.

\begin{figure}[th]
\includegraphics[width=8cm]{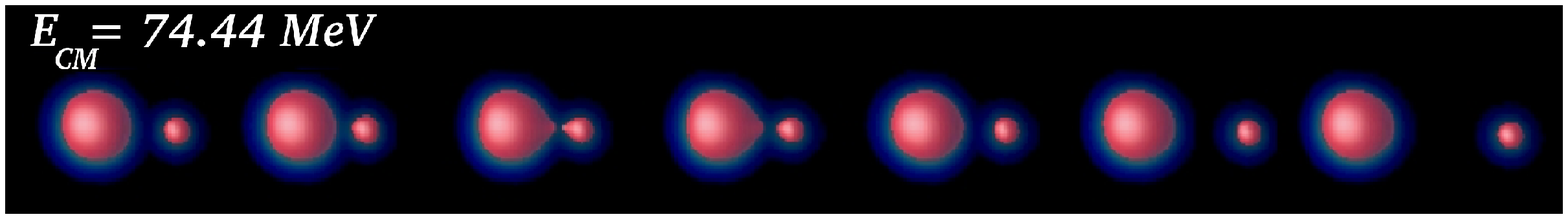}
\hfill
\includegraphics[width=8cm]{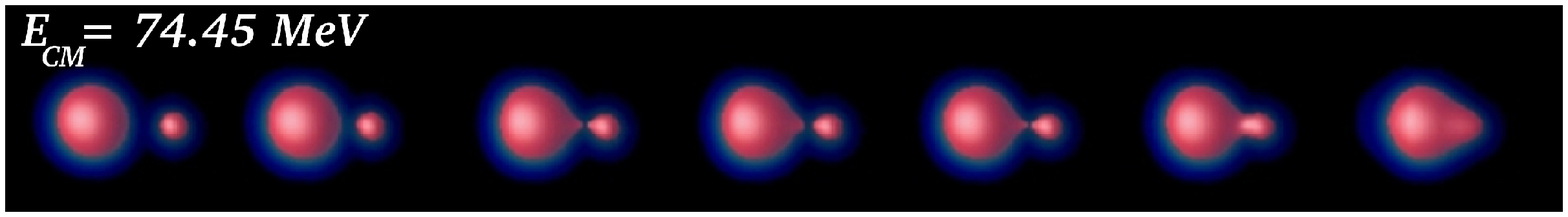}
\caption{Densities associated to a $^{16}$O+$^{208}$Pb 
central collision at a center of mass energy $E=74.44$~MeV (top) and
74.45~MeV (bottom).
The surface corresponds to an isodensity at half the saturation density. 
Density decreases from red to black.
Each plot is separated by 135 fm/c.}
\label{fig:74_44}
\end{figure}

Figure~\ref{fig:74_44} shows the density plot for a $^{16}$O+$^{208}$Pb 
central collision at a center of mass energy $E_{CM}=74.44$~MeV (top) and $E_{CM}=74.45$~MeV (bottom). 
We see that the system separates into two fragments after a neck formation at the lowest energy. However, adding 10~keV is enough to fuse. We deduce 
the fusion barrier $V_B^{TDHF}= 74.445\pm0.005$~MeV for this system.
This value is in excellent agreement with 
the experimental one
$V_B^{exp}\simeq 74.5$~MeV~\cite{mor99}. 

\begin{figure}[th]
\includegraphics[width=8cm]{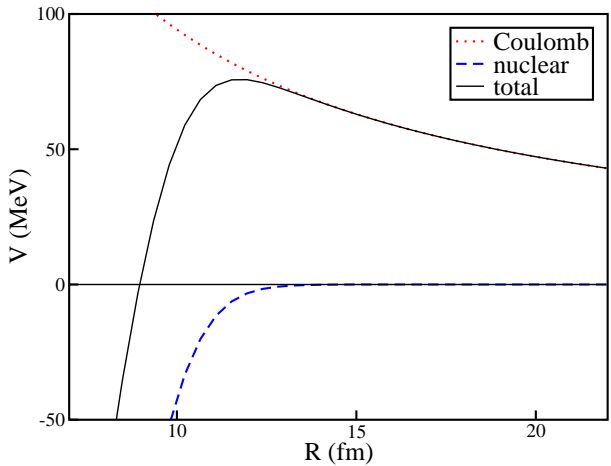}
\caption{Nucleus-nucleus potential of $^{16}$O+$^{208}$Pb assuming frozen HF densities of the nuclei  at various internuclear distances $R$.}
\label{fig:frozen}
\end{figure}

To see if these results are affected by dynamical effects,
we compare them with a purely static approach. 
Figure~\ref{fig:frozen} shows the nucleus-nucleus potential of this system 
if we assume frozen HF densities at various internuclear distances $R$ using the same EDF and the same numerical approximations as in the TDHF calculations. Here, repulsion due to Pauli principle is neglected, which
 is a rather good approximation for distances down to the barrier as overlap of the nuclear densities are small.
With this approach, the barrier is $V_B^{frozen}=76$~MeV~\cite{was08}.
This is slightly higher than the TDHF value. 
This difference is due to dynamical
effects such as transfer (see Section~\ref{sec:transfer}) that reduce the barrier.

\begin{figure}[th]
\includegraphics[width=8cm]{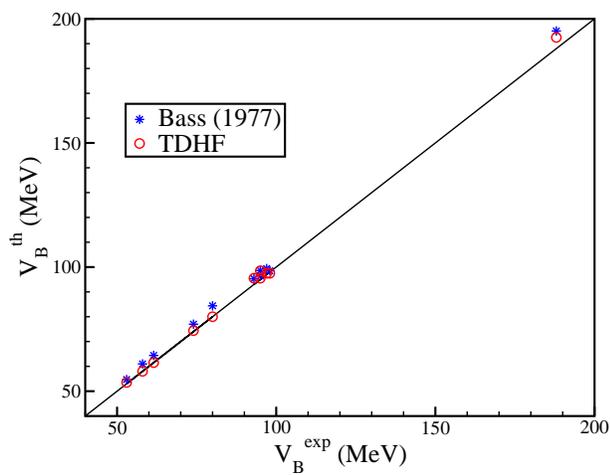}
\caption{Theoretical fusion barriers from TDHF calculations (circles) 
and the Bass barriers (stars) as function of the experimental values
(from  barrier distributions centroid).}
\label{fig:barriers}
\end{figure}

Figure~\ref{fig:barriers} shows a comparison between experimental 
fusion barriers and those from TDHF calculations
for systems with various total masses and mass asymmetries.
The lowest barrier is for $^{40}$Ca+$^{40}$Ca and the highest one for 
$^{48}$Ti+$^{208}$Pb.   
The  agreement with experimental data is slightly better with TDHF than with the Bass barrier~\cite{bas77}.
Remembering that these TDHF calculations have no adjustable parameter on reaction mechanisms, 
we conclude that one can use it for fusion barriers predictions with confidence.

\subsection{Transfer below the barrier in $^{16}$O+$^{208}$Pb}
\label{sec:transfer}

It has been suggested that transfer may affect the barrier 
in the $^{16}$O+$^{208}$Pb system~\cite{tho89}.
Indeed, we see in Fig.~\ref{fig:74_44} 
that the two fragments are linked 
by a neck and form a dinuclear system during $\sim400$~fm/c. 
It is enough time for the nuclei to transfer nucleons,
leading to a dynamical evolution of the barrier.  

\begin{figure}[th]
\includegraphics[width=8cm]{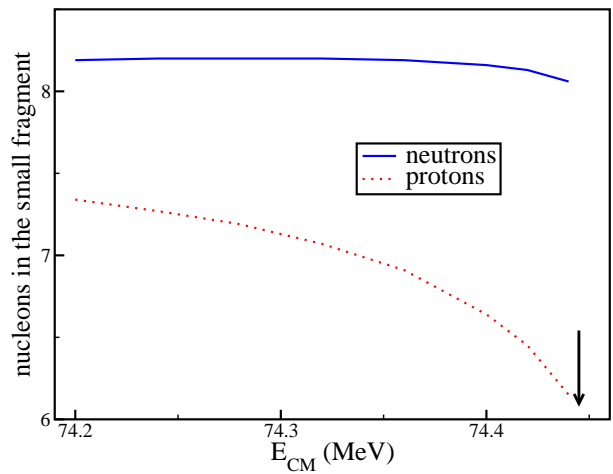}
\caption{Transfer below the barrier in $^{16}$O+$^{208}$Pb. 
The average number of protons and neutrons of the small fragment
are plotted as function of energy. The arrow indicates the TDHF fusion barrier.}
\label{fig:transfer}
\end{figure}

To get a deeper insight into the transfer preceding fusion in this reaction, we have plotted the evolution of the numbers of protons and neutrons of the small fragment as function of energy in Figure~\ref{fig:transfer}.
Almost two protons and no neutron, in average, 
have been transfered from the $^{16}$O to the $^{208}$Pb. 
The two-proton transfer from the light to the heavy nucleus 
 is then expected to be an important channel at the barrier. 
 This is consistent with experimental observations of 
 relatively high Carbon production cross sections, 
 of the same order of the Nitrogen ones, in the exit channel
 of $^{16}$O+$^{208}$Pb at the barrier~\cite{vid77,vul86}.
Note that we have to distinguish between the transfer of typically
 one or two nucleons and more violent collisions like deep-inelastic reactions.
 In the latter, the width of the particle number distribution is known to be 
 underestimated with a single Slater determinant~\cite{das79}.

\subsection{Fusion cross section of $^{16}$O+$^{208}$Pb}

\begin{figure}[th]
\includegraphics[width=8cm]{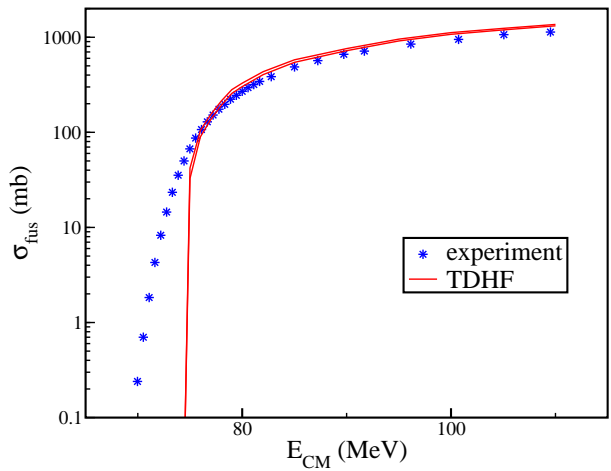}
\caption{Experimental fusion excitation 
function (stars) for $^{16}$O+$^{208}$Pb from Ref.~\cite{mor99}. The lines denote the upper and lower
limits for the fusion cross sections obtained from TDHF calculations.}
\label{fig:fus}
\end{figure}

We now focus on fusion cross sections. In TDHF, the latter can be approximated by
$\si_{fus}(E_{CM}) \simeq \pi b_c^2$ where $b_c$ is the critical impact parameter associated to the initial Coulomb trajectrory below which fusion occurs~\cite{sim08b}.
Figure \ref{fig:fus} shows the excitation function obtained for the 
$^{16}$O+$^{208}$Pb system in comparison to experimental data~\cite{mor99}.
There is a good agreement above the barrier, though
the fusion cross sections are overestimated by about 16~$\%$.
The origin of this small discrepancy is nonetheless puzzling as the TDHF fusion barrier agrees perfectly with experiment. More precise investigations  of fusion above the barrier with TDHF are then needed. 

Below the barrier, the TDHF fusion cross section vanishes, following 
a classical behavior. This drawback of TDHF is well known and is 
due to the restriction to a single independent particles state. 
Indeed, to get a fusion probability
between 0 and 1, we need at least two Slater determinants: 
one describing the two well separated fragments after the collision
when fusion does not occur
and one describing the fused system. 
It is necessary to go beyond TDHF to treat a sum of Slater determinants
and then to describe 
sub-barrier fusion due to quantum tunneling.

\section{Fusion hindrance of quasisymmetric heavy nuclei}
\label{sec:fus_heavy}

As we showed in the preceding section, TDHF systematics are in good agreements 
with experimental fusion barriers for light or medium-heavy systems. In these cases,
models where the nuclei are assumed to be frozen with their ground state density give already a good approximation to fusion barriers.
However, when collision partners get heavy and symmetric, 
typically when the product of their charges is such that $Z_1 \times Z_2 > 1600$, 
such frozen approaches significantly underestimate fusion barriers~\cite{gag84}.
These systems need an extra energy to fuse compared to standard barrier height estimations.  This phenomenon gave birth to the {\it extra-push} model~\cite{swi82}. 
One-body dissipation of the kinetic energy towards nucleonic degrees of 
freedom is often invoked to be responsible for this fusion hindrance. Thus, TDHF 
theory, which incorporates such one-body dissipation in the microscopic description of nuclear dynamics, should be appropriate to study fusion mechanisms in heavy (quasi)-symmetric systems.

\begin{figure}[th]
\includegraphics[width=8cm]{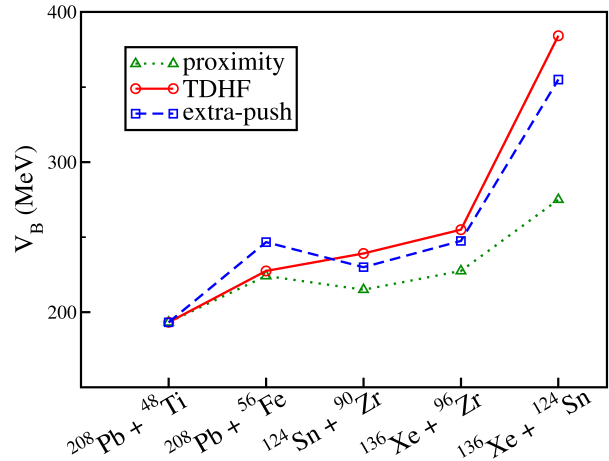}
\caption{Fusion barriers obtained for various systems with the proximity model~\cite{blo77} (dotted line),
the extra-push model~\cite{swi82} (dashed line) and TDHF (solid line).}
\label{systematics:xP}
\end{figure}

Let us now present a preliminary study of $^{48}$Ti+$^{208}$Pb, 
$^{56}$Fe+$^{208}$Pb, 
$^{90}$Zr+$^{124}$Sn, $^{96}$Zr+$^{136}$Xe and $^{124}$Sn+$^{136}$Xe head-on collisions. Due to computational time limitations, 
we assume that the nuclei have fused if a compact system is formed for a minimum duration of 1200 fm/c. 
Moreover, HF ground statse of $^{124}$Sn and $^{136}$Xe are slightly deformed. 
As a first approximation, we neglect possible effects of deformation on fusion 
and consider only one orientation to limit computational effort. 

Results are shown on Fig.~\ref{systematics:xP}, and compared with 
the extra-push model~\cite{swi82} and the proximity predictions of fusion 
barriers~\cite{blo77}. 
 As we can see, TDHF results are in rather good agreement with 
the extra-push model. Both approaches predict significantly higher fusion barriers than the proximity model when mass symmetry increases.
 TDHF should then be useful tool to investigate dissipative mechanisms responsible for fusion hindrance. 

\begin{figure}[th]
\includegraphics[width=8cm]{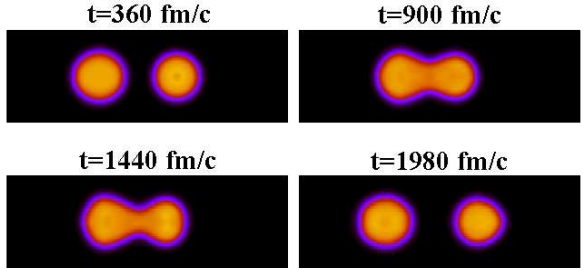}
\caption{Snapshots of the density profile in the reaction plane of 
the $^{90}$Zr+$^{124}$Sn central collision at $E_{CM}=235$~MeV.
Hot colors are associated
to densities close to the saturation density $\rho_0=0.16$~fm$^{-3}$. }
\label{density_profile:124Sn90Zr_235CoM}
\end{figure}

 Fusion models usually rely on the reduction of the complex dynamical many-body problem into few relevant degrees of freedom, like neck parameters, mass asymmetry or distance between fragments. 
 The TDHF theory, however, is free of such choice of collective variables. 
Thus, the shape of the system can evolve freely in three dimensions. For example,
we show on Fig.~\ref{density_profile:124Sn90Zr_235CoM} snapshots of the density profile in the reaction plane of 
the $^{90}$Zr+$^{124}$Sn central collision at $E_{CM}=235$~MeV.
This bombarding energy is $17$~MeV higher than 
the barrier given by the Hartree-Fock frozen density approximation ($V_B^{frozen}=218$~MeV) and $3.75$~MeV 
below the TDHF fusion threshold ($V_B^{TDHF}=238.75$~MeV). The separation of the system in 
two fragments is thus entirely driven by the fusion hindrance mechanisms. 

We observe in Figure~\ref{density_profile:124Sn90Zr_235CoM} the formation of a neck with a density close to saturation density. 
Also, this neck survives during approximately $1000$~fm/c. This time is long enough to allow for a strong rearrangement of the system via nucleon transfer and dynamical shape evolution which modifies the fusion potential landscape. However, we also observe in Fig.~\ref{density_profile:124Sn90Zr_235CoM} that this rearrangement is limited as it does not allow for the two fragments to merge and form a single one, i.e., the system keeps clearly the shape of a dinuclear system. 
These observations are encouraging and studies of isospin equilibration, mass transfer and shape evolution during the contact configuration are ongoing.

\section{Collision time of actinides}
\label{sec:U+U}

Collisions of two Uranium or heavier nuclei are used, amongst other things, to investigate the physics of super-strong electric fields~\cite{gre83}. 
Indeed, it has been proposed that a giant system with total charge $Z \geq 173$ may induce spontaneous electron-positron pair emissions from vacuum by a fundamental quantum electrodynamics (QED) process~\cite{rei81,gre83}. 
However, no experimental evidence of this process has been obtained so far~\cite{ahm99}.
The keypoint might be a good understanding of collision dynamics, in particular a reliabe prediction of collision times.
In the $^{238}$U+$^{238}$U reaction, one might be able to see this process for sticking times of several~$10^{-21}$~s as shown by recent theoretical calculations based on the time-dependent Dirac equation~\cite{ack08}.

\begin{figure}[th]
\includegraphics[width=8cm]{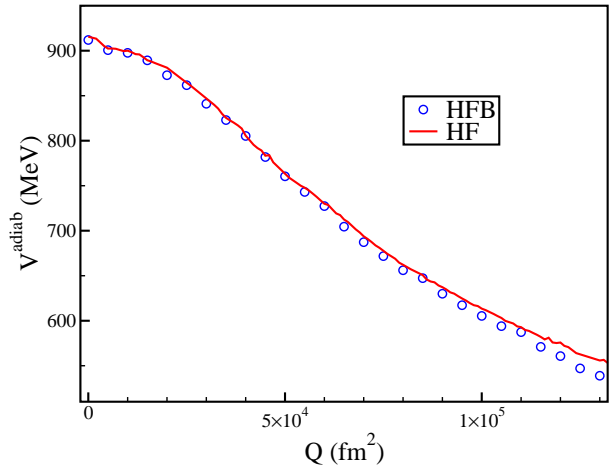}
\caption{Symmetric fission valley of the $^{238}$U+$^{238}$U giant system. 
The Hartree-Fock (solid line) and Hartree-Fock-Bogoliubov (circles) adiabatic potential energies under a quadrupole constraint are plotted as function of the total quadrupole moment.}
\label{fig:adiab}
\end{figure}

Figure~\ref{fig:adiab} shows that there is no pocket in the nucleus-nucleus potential of the $^{238}$U+$^{238}$U system, in agreement with Refs.~\cite{ber90,tia08}.
However, dissipation mechanisms such as evolution of nuclear shapes may delay the separation of the system~\cite{zag06}. 
Experimentally, collision times have been estimated from oscillations in $\delta-$electron spectra~\cite{sof79} and values of $\sim2\times10^{-21}$~s have been obtained in $^{238}$U+$^{238}$U above the barrier~\cite{kri86}.
Recently, delay times in this reaction was searched analyzing kinetic energy loss and mass transfer~\cite{gol08}.
On the theoretical side, macroscopic models 
have first been used (see, e.g., Ref.~\cite{zag06}).
However, the complexity of reaction mechanisms and the high number of degrees of freedom to be included in realistic calculations motivate the use of microscopic approaches. First microscopic calculations of the collision of two $^{238}$U nuclei have been performed recently thanks to the Quantum Molecular Dynamics (QMD) model~\cite{tia08}. 

\begin{figure}[th]
\includegraphics[width=8cm]{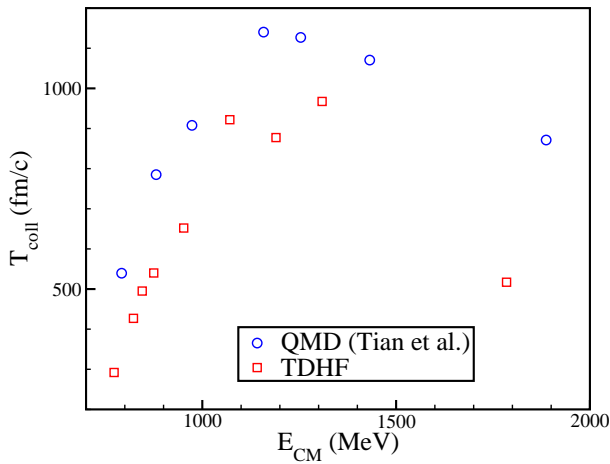}
\caption{Collision time of two $^{238}$U nuclei as function of center of mass energy.}
\label{fig:Tcoll_Ecm}
\end{figure}

Here, we report preliminary results on central collisions of two $^{238}$U with deformation axis parallel to each other and perpendicular to the collision axis. 
The collision time $T_{coll}$ between nuclei  is defined, here, as the time during which the neck density exceeds $\rho_0/10=0.016$~fm$^{-3}$. 
Figure~\ref{fig:Tcoll_Ecm} shows the evolution of  $T_{coll}$ as a function of $E_{CM}$.
We see a rise and fall of $T_{coll}$ with a maximum of $\sim 1000$~fm/$c$ at $1000\leq E_{CM}\leq1300$~MeV, in a rather good agreement with the QMD calculations of Ref.~\cite{tia08}. 
Central collisions in this energy range are then expected to produce superstrong electric fields with lifetimes which fill the necessary condition to observe $e^+e^-$ pair excitations from QED vacuum.

\section{Conclusions and Perspectives}

We presented some applications of the TDHF
theory to heavy ion collisions around the barrier. The only phenomenological input 
is the set of parameters of the energy density functional which have not been
adjusted on any reaction mechanism.
Despite this, the agreement between the TDHF calculations 
and the experimental fusion barriers is excellent for a wide range 
of projectiles and targets. However, these calculations overestimate
 the fusion cross sections for the system $^{16}$O+$^{208}$Pb 
 above the barrier by about 16$\%$. Theoretical investigations are needed to understand this (rather small) disagreement.

Though the TDHF fusion cross section has a classical behavior, 
 the quantum nature of the single-particle 
wave functions is well treated. Thus, we presented a study
of transfer in the $^{16}$O+$^{208}$Pb below the barrier. 
Two-proton transfer from the $^{16}$O to the $^{208}$Pb nucleus is dominant at the barrier. This qualitative observation is in agreement with experiment.
 
We are presently extending the fusion study to heavy quasi-symmetric systems to investigate the phenomena responsible for the fusion hindrance in such systems. 
Preliminary results are encouraging as they indicate that this fusion hindrance is well reproduced by TDHF.

We finally presented calculations of collision times between two $^{238}$U nuclei.
Our conclusion is that the giant system might survive enough time ($\sim1000$~fm/$c$ at an energy $1000\leq E_{CM}\leq1300$~MeV) to test nonperturbative quantum electrodynamics with super-strong electric fields.

\section*{Acknowledgements}

We thank P.~Bonche who provided his code. 
We are  grateful to M.~Dasgupta, R.~Dayras, A.~Drouart, W.~Greiner, D.~Hinde, D.~Lacroix, W.~Mittig, A.~Villari and K.~Washiyama for fruitful discussions. 
The calculations have been performed in the 
Centre de Calcul Recherche et Technologie of the Commisariat \`a l'\'Energie Atomique.

\end{document}